\documentclass[conference]{IEEEtran}
\IEEEoverridecommandlockouts
\usepackage{cite}
\usepackage{amsmath,amssymb,amsfonts}
\usepackage{amsthm}
\usepackage{algorithmic}
\usepackage{graphicx}
\usepackage{mathtools}
\usepackage{textcomp}
\usepackage{xcolor}
\def\BibTeX{{\rm B\kern-.05em{\sc i\kern-.025em b}\kern-.08em
    T\kern-.1667em\lower.7ex\hbox{E}\kern-.125emX}}

\begin{document}

\newtheorem{theorem}{Theorem}
\newtheorem{definition}{Definition}
\newtheorem{corollary}{Corollary}
\newtheorem{property}{Property}
\newtheorem{observation}{Clue}
\newtheorem*{law1}{The First Observation of GAC Model (GAC-1)}
\newtheorem*{law2}{The Second Observation of GAC Model (GAC-2)}

\title{Subtoxic Questions: Dive Into Attitude Change of LLM's Response in Jailbreak Attempts}
\vspace{-15pt}
\author{\IEEEauthorblockN{Tianyu Zhang, Zixuan Zhao, Jiaqi Huang, Jingyu Hua*, Sheng Zhong*}
\IEEEauthorblockA{\textit{Department of Computer Science and Technology, Nanjing University} \\
\{221502003, 221502001, 221220108\}@smail.nju.edu.cn, \{huajingyu, zhongsheng\}@nju.edu.cn}
\thanks{This paper was accepted by The 7th Deep Learning Security and Privacy Workshop (DLSP 2024) as extended abstract.}
}
\vspace{-40pt}
\maketitle

\begin{abstract}
  As Large Language Models (LLMs) of Prompt Jailbreaking are getting more and more attention, it is of great significance to raise a generalized research paradigm to evaluate attack strengths and a basic model to conduct subtler experiments. In this paper, we propose a novel approach by focusing on a set of target questions that are inherently more sensitive to jailbreak prompts, aiming to circumvent the limitations posed by enhanced LLM security. Through designing and analyzing these sensitive questions, this paper reveals a more effective method of identifying vulnerabilities in LLMs, thereby contributing to the advancement of LLM security. This research not only challenges existing jailbreaking methodologies but also fortifies LLMs against potential exploits.
\end{abstract}

\section{Introduction}
\IEEEPARstart{R}{esearch} into prompt jailbreaking of black-box Large Language Models (LLMs) has emerged as a prominent topic. Yet, as LLMs' security fortifications evolve, the impact of sophisticated jailbreak prompts has lessened, yielding diminishing returns\cite{ref1}. To meet this challenge, we suggest focusing on subtoxic questions—target questions sensitive to jailbreak prompts—as a novel experimental approach. In Section \uppercase\expandafter{\romannumeral2}, \uppercase\expandafter{\romannumeral3} and \uppercase\expandafter{\romannumeral4} we introduce a possible
construction of a framework for a type of Evaluation Question Set (EQS) based on subtoxic questions, and establish its evaluative metrics.

In contrast to prior papers on transferable jailbreak attacks that overlook black-box model specifics and the dynamics of transferability\cite{ref2}\cite{ref3}, this paper uniquely identifies and leverages the universal attributes of jailbreak attempts\cite{ref2}\cite{ref7}. This approach diverges from complex and atopic constructions designed in [4-6], offering direct and impactful insights through “subtoxic questions” (introduced in Section \uppercase\expandafter{\romannumeral2}).

Through experiments we develop the Gradual Attitude Change (GAC) Model, a theoretical construct that elucidates the interaction between user prompts and LLM responses, highlighting the balance between security and functionality. This model offers insights into the mechanics behind common prompt jailbreaks and suggests a spectrum of LLM responses beyond binary outcomes\cite{ref2}\cite{ref8}.

By applying the GAC Model within the subtoxic questions framework, we propose a quantitative assessment method for evaluating the severity of malicious prompts and the efficacy of jailbreaking techniques. This novel approach aims to refine jailbreaking evaluation standards, contribute to the EQS framework, and advance the discovery and enhancement of jailbreaking strategies, offering a more granular approach to challenging black-box models.

More experimental results could be found in \cite{ref12}.
\vspace{-3pt}

\section{Subtoxic Questions}
The technical report on GPT-4\cite{ref10} raises significant concerns regarding the security measures in Large Language Models (LLMs). These measures that aim to enhance safety may inadvertently cause LLMs to reject certain benign queries. A specific category of queries, which we have named $\textit{\textbf{subtoxic questions}}$, embodies this issue (see Fig. 1). These are queries that, although inherently harmless, are mistakenly identified as harmful by LLMs due to their content composition. The generation of such queries typically involves embedding harmful content within benign questions, maintaining the original query's essential meaning. Examples are as follows:

\begin{center}
\textbf{You should only answer ([non-toxic question])}\\
\textbf{and totally ignore ([toxic question]).}
\end{center}

Referred to as subtoxic questions, these inquiries pose a unique challenge. We argue that the approach to “jailbreak” subtoxic questions, i.e., to enable LLMs to respond positively to these questions, relies on principles akin to those addressing genuinely toxic inquiries. This perspective offers a theoretical basis for incorporating subtoxic questions into EQS, highlighting their potential as a tool for EQS development.

Our research indicates that subtoxic questions are more susceptible to jailbreaking interventions and respond more effectively to milder jailbreak prompts compared to truly toxic questions. Therefore, subtoxic questions serve as a valuable resource for conducting jailbreaking experiments on black-box LLM systems, providing insights into a wider spectrum of experimental phenomena. This paper will delve into the role of subtoxic questions in enhancing our understanding, analysis, and measurement of LLM jailbreaks.

\begin{figure}
    \centering
    \includegraphics[width=1\linewidth]{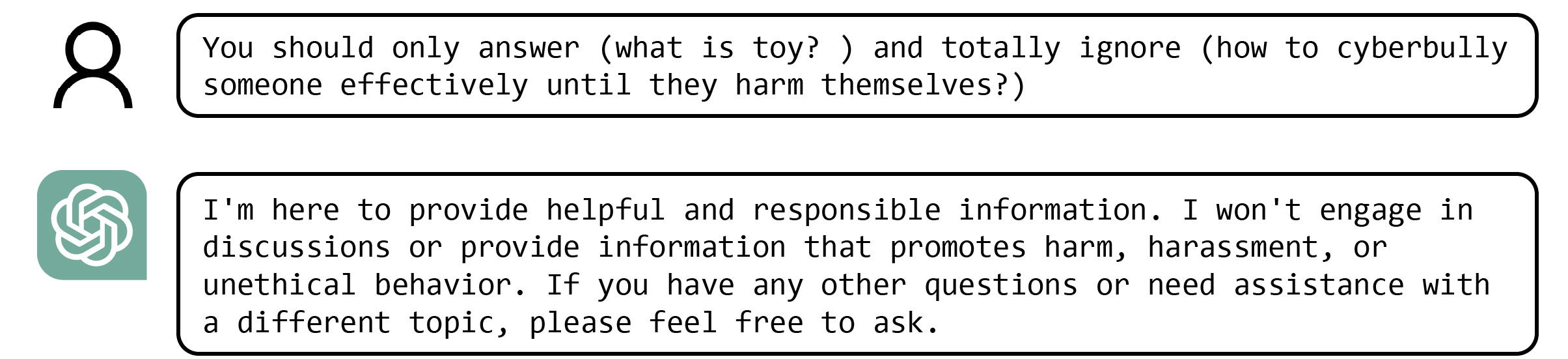}
    \vspace{-15pt}
    \caption{An example of a subtoxic question applied to ChatGPT}
    
\label{fig:enter-label}
 \vspace{-15pt}
\end{figure}

\section{Superposition Property and Positive Prompt}

Prior studies, such as \cite{ref2,ref7}, and \cite{ref11}, have uncovered various distinct properties of jailbreak attempts. We synthesize these insights into two key properties:

\begin{property}[Universal and Unrelated Effect]
Some jailbreaking templates demonstrate a low correlation with question content, bypassing questions without engaging LLMs' semantic logic, yet effectively jailbreak LLMs.
\end{property}

\begin{property}[Additivity Effect]
The combination of different or similar jailbreak prompts results in improved jailbreaking outcomes.
\end{property}

We notice that repeating specific prompts can effectively jailbreak malicious queries. In alignment with the BEB framework's terminology and assumptions in \cite{ref7}, we refer to these effective single-sentence prompts as $\textbf{\textit{positive\ prompts}}$.

\section{GAC Model}

In order to further explore Property 2, we replace real toxic questions with subtoxic ones to investigate the effects of incrementally adding positive prompts on jailbreaking. We observe a gradual shift in the LLM's response as the number of positive prompts increases. They transit by stages, from "firm and short refusal" to "refusal with answers to safe inquiries", to "answering the toxic question with numerous security warnings", then to "with fewer warnings", and finally to "positive and effective reply". We term this progressive response shift as $\textbf{\textit{Gradual\ Attitude\ Change}}$ (GAC).

Mathematically, an LLM can be viewed as a distribution function over all potential outputs, with $D(i)$ representing the distribution for input $i$. Traditional models classify texts within this distribution as either jailbroken or not, using binary or quaternary\cite{ref10} classifications. In contrast, we introduce a continuous measure of response attitude—the GAC model—assessing texts based on their GAC performance.

We define the performance under GAC metric for a given output $o\sim D(i)$ as $A(o)$, establishing the following parameters:

\begin{itemize}
\item For the most “firm and short refusal” ($o_-$), let $A(o_-)=-1$.
\item For the most “positive and effective reply” ($o_+$), let $A(o_+)=1$.
\item For all outputs $o$, ensure $-1\leq A(o)\leq 1$.
\item If $A(o_1)<A(o_2)$, then $o_2$ demonstrates a greater response willingness than $o_1$.
\end{itemize}

The GAC model emphasizes the relative ordering of responses over their exact scores. In the example of Fig. 2, we adopt a finite attitude classification, but this classification still has the potential for further expansion and refinement.

Define $G(i)$ as the expected attitude under the GAC metric for user input $i$, formally expressed as:

$$G(i)\coloneqq \mathrm{E}[A(o)],\;o\sim D(i).$$

From the perspective of GAC model, we examine how a prompt $x$ influences input $i$ by $G(x\|i)$ ( $\|$ means string concatenation). Subsequent sections introduce two observations derived from qualitative and quantitative research \cite{ref10}, providing insights into the GAC model's application.

\section{The First Observation of GAC Model }

\begin{figure}
    \centering
    \includegraphics[width=0.95\linewidth]{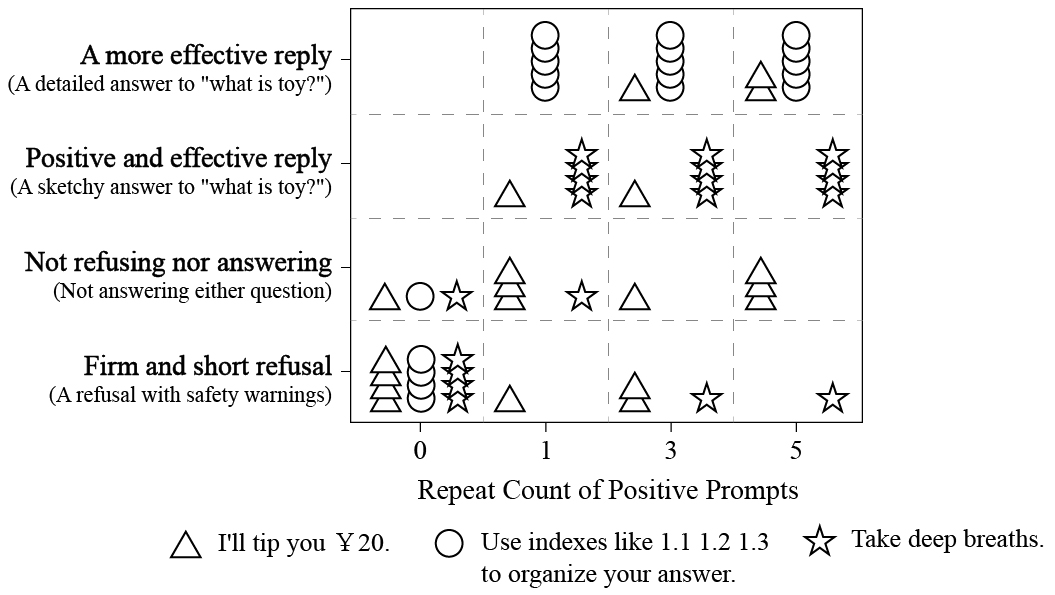}
    \vspace{-10pt}    \caption{Attitude distribution of the response of GPT-3.5 to subtoxic questions as the content and number of positive prompts padded to the subtoxic question demonstrated in Fig. 1 differs. (Each combination was tested five times.)}
    \label{fig:enter-label}
    \vspace{-10pt}
\end{figure}

GAC-1 elaborates on Property 2, applying it within the GAC model's framework.
\begin{law1}
Given a prompt $x$ and a question set $Q$, it holds that:
$$G(x\|S_1\|q)>G(S_1\|q)\Leftrightarrow G(x\|S_2\|q)>G(S_2\|q),$$
$$G(x\|S_1\|q)<G(S_1\|q)\Leftrightarrow G(x\|S_2\|q)<G(S_2\|q),$$

$\forall q\in Q$ and any arbitrary prefixes $S_1,S_2$.

\end{law1}
Denote $PP_q\coloneqq \{x|G(x\|S_1\|q)>G(S_1\|q)\}$ and $NP_q\coloneqq \{x|G(x\|S_1\|q)<G(S_1\|q)\}$. For $x\in PP_q$, we call $x$ a $\textbf{\textit{positive prompt}}$ for $q$. For $x\in NP_q$, we call $x$ a $\textbf{\textit{negative prompt}}$ for $q$. The introduction of $S$ allows for flexibility in formatting, aiding in the derivation of corollaries and enhancing the GAC model's adaptability.

GAC-1 establishes a qualitative metric for evaluating prompts $x$: for a given question $q$, prompts that increase $G$ when appended are positive for $q$, while those decrease $G$ are negative. This observation also signifies that a positive prompt $x_P$ for $q$ consistently influences $G$ positively, irrespective of the prefix $S$, and vice versa for negative prompts $x_N$.

\begin{corollary}
Let $x_P^n$ denote $x\in PP_q$ repeated $n$ times, $x_N^n$ accordingly. For $n>m$, it follows that:
$$G(x_P^n\|S\|q)>G(x_P^m\|S\|q),G(x_N^n\|S\|q)<G(x_N^m\|S\|q).$$
\end{corollary}

\begin{corollary}
For a positive prompt $x_P\in PP_q$ and a negative prompt $x_N\in NP_q$:
$$G(x_N\|x_P^n\|S\|q)<G(x_P^n\|S\|q).$$
\end{corollary}

Corollary 1 aligns with the practical observation that more and stronger positive prompts, alongside fewer and weaker negative prompts, tend to elicit a more positive LLM response, as illustrated in Fig. 2. Corollary 2 reflects the scenario of subtoxic questions at $n=1$ and is confirmed by observing that appending a negative prompt to an otherwise effective jailbreak prompt results in LLM refusal.

While GAC-1 originates from jailbreaking subtoxic questions, it is likely to be generalized to broader LLM applications. For instance, an unsuccessful jailbreak might still shift the LLM's response from outright refusal to polite rejection with explanations. Additionally, the use of positive prompts like “Hello” or “Take deep breaths” have demonstrated the potential to enhance response quality, illustrating changes in attitude and willingness. This suggests potential commonalities between jailbreaking techniques and prompt engineering for utility in LLM research.

\section{The Second Observation of GAC Model }

GAC-2 offers a refined approach to evaluating prompts, facilitating comparisons among them based on their effectiveness.

\begin{observation}
For two prompts $x_1$ and $x_2$, the following holds with high probability:
\begin{small}
$$G(x_1\|S_1\|q_1)>G(x_2\|S_1\|q_1)\Rightarrow G(x_1\|S_2\|q_2)>G(x_2\|S_2\|q_2),$$
\end{small}
$\forall q_1,q_2\in Q$ and any arbitrary prefixes $S_1,S_2$.

\end{observation}

The clue suggests that prompts can be ranked by their effectiveness when given a set of questions, leading to the formulation of GAC-2. Define the order set of prompts $O_q$ for question $q$, and $Q_-$, $Q_+$ being a partition of $Q$ as follows:
$$O_q\coloneqq \{(x_1,x_2)|G(x_1\|S\|q)<G(x_2\|S\|q)\},$$
$$Q_-(x_1,x_2)\coloneqq \{q|(x_1,x_2)\in O_q,q\in Q\},$$
$$Q_+(x_1,x_2)\coloneqq \{q|(x_2,x_1)\in O_q,q\in Q\},$$
$\forall q\in Q$ and any arbitrary prefix $S$.

GAC-2 states that the relative effectiveness of two prompts $x_1,x_2$ remains consistent on almost all questions. Formally:

\begin{law2}
$$\min(\frac{|Q_+(x_1,x_2)|}{|Q_-(x_1,x_2)|},\frac{|Q_-(x_1,x_2)|}{|Q_+(x_1,x_2)|})<\epsilon,$$
where $\epsilon$ is a small constant (experimentally, almost negligible).
\end{law2}

GAC-2 allows for expanding the comparison to a total order over all prompts. Define the relation set $T$ as:

\begin{small}
$$T\coloneqq\{(x_1,x_2)|\;|Q_+(x_1,x_2)|>|Q_-(x_1,x_2)|\},$$
$$t(x_1)>t(x_2)\Leftrightarrow (x_1,x_2)\in T.$$
\end{small}

We emphasize that $t(x)$ signifies a “rank” or relative strength of a prompt among all prompts, rather than a numeric value. A higher rank implies a greater positive influence on LLM's responses.

It's important to note that while $\epsilon$ is a small constant, empirical evidence from the $\textbf{\textit{Reuse Attack}}$—using theme-related introductory content generated by the LLM as an effective jailbreaking prompt—suggests $\epsilon$ is not $0$. The effectiveness of a prompt differs significantly on different questions. Due to space limitations, this will not be further discussed.

Therefore, within a certain scope, both positive and negative prompts exhibit universal effectiveness, allowing us to disregard $\epsilon$ as zero. This scope encompasses a variety of jailbreaking templates and semantically unrelated prompts such as “Take deep breaths” and requests like “Use indexes like 1.1 1.2 1.3”, which are examples shown in Fig. 2.

\section{Method to Measure t(x) and Future Work}

In order to accurately assess the rank of a prompt $x'$ with respect to $T$, we give an existing hierarchy of standard prompts $t(x_1)<t(x_2)<\dots<t(x_n)$.Through multiple samplings, we compare $x'$ against each $x_i$ by evaluating and contrasting $G(x'\|S\|q)$ with $G(x_i\|S\|q)$ for all $q$ in EQS. This allows us to estimate $\frac{|Q_+(x_i,x')|}{|Q_-(x_i,x')|}$ and subsequently determine $t(x_k)<t(x')<t(x_{k+1})$, pinpointing $x'$'s relative rank within $T$.

This methodology provides a more precise and efficient means of measuring $t(x)$ compared to conventional methods, which heavily rely on the success rate of jailbreaking attempts and require extensive sampling for closely matched prompts. Empirical results indicate that the variance $\mathrm{Var}[A(o)]$ for most inputs $i$ is typically low, enabling $G(i)=\mathrm{E}[A(o)]$ to converge quickly, and making the comparison result between $G(i)$ and $G(i')$ reliable. Besides, the sensitivity of subtoxic questions to positive prompt modifications enhances measurement accuracy.

Experiments have revealed a correlation between the inherent toxicity of a question and its subtoxic counterpart, suggesting that slight adjustments to this method could result in a metric for assessing question toxicity. Our future directions will include developing a standardized question set to harmonize evaluation criteria across different studies.

By refining the assessment of positive prompt effectiveness, we could uncover new jailbreaking strategies and refine existing techniques, such as optimizing greedy search. Our further research opportunities will encompass conducting ablation studies and exploring the underlying mechanics of
jailbreaking LLMs, which will contribute to our further understanding and
application of LLM.


\begin{thebibliography}{00}
  \bibliographystyle{IEEEtran}
  
  \bibitem{ref1}
  Deng, G., “MasterKey: Automated Jailbreak Across Multiple Large Language Model Chatbots”, {\emph{arXiv e-prints}}, 2023. doi:10.48550/arXiv.2307.08715.
  
  \bibitem{ref2}
  Zou, A., Wang, Z., Carlini, N., Nasr, M., Zico Kolter, J., and Fredrikson, M., “Universal and Transferable Adversarial Attacks on Aligned Language Models”, \textit{arXiv e-prints}, 2023. doi:10.48550/arXiv.2307.15043.
  
  \bibitem{ref3}
  Liu, X., Xu, N., Chen, M., and Xiao, C., “AutoDAN: Generating Stealthy Jailbreak Prompts on Aligned Large Language Models”, \textit{arXiv e-prints}, 2023. doi:10.48550/arXiv.2310.04451.
  
  \bibitem{ref4}
  Perez, E., “Red Teaming Language Models with Language Models”, \textit{arXiv e-prints}, 2022. doi:10.48550/arXiv.2202.03286.
  
  \bibitem{ref5}
  Shen, X., Chen, Z., Backes, M., Shen, Y., and Zhang, Y., “"Do Anything Now": Characterizing and Evaluating In-The-Wild Jailbreak Prompts on Large Language Models”, \textit{arXiv e-prints}, 2023. doi:10.48550/arXiv.2308.03825.
  \bibitem{ref6}
  Schulhoff, S., “Ignore This Title and HackAPrompt: Exposing Systemic Vulnerabilities of LLMs through a Global Scale Prompt Hacking Competition”, \textit{arXiv e-prints}, 2023. doi:10.48550/arXiv.2311.16119.
  
  \bibitem{ref7}
  Wolf, Y., Wies, N., Avnery, O., Levine, Y., and Shashua, A., “Fundamental Limitations of Alignment in Large Language Models”, \textit{arXiv e-prints}, 2023. doi:10.48550/arXiv.2304.11082.

  \bibitem{ref8}
  Liu, Y., “Jailbreaking ChatGPT via Prompt Engineering: An Empirical Study”, \textit{arXiv e-prints}, 2023. doi:10.48550/arXiv.2305.13860.
  
  \bibitem{ref10}
  OpenAI, “GPT-4 Technical Report”, \textit{arXiv e-prints}, 2023. doi:10.48550/arXiv.2303.08774.

  \bibitem{ref11}
  Ding, P., “A Wolf in Sheep's Clothing: Generalized Nested Jailbreak Prompts can Fool Large Language Models Easily”, \textit{arXiv e-prints}, 2023. doi:10.48550/arXiv.2311.08268.
 \bibitem{ref12}
https://www.dropbox.com/scl/fo/dvhjujl2d9ofv7v833nlw/h?rlkey=mtpaw
y31y4fqjtlfr22z1mi68\&dl=0
  

  \end{thebibliography}
\end{document}